\documentclass{article}

\usepackage{amsmath,amsfonts}
\usepackage{algorithmic}
\usepackage{algorithm}
\usepackage{array}
\usepackage[caption=false,font=normalsize,labelfont=sf,textfont=sf]{subfig}
\usepackage{textcomp}
\usepackage{stfloats}
\usepackage{url}
\usepackage{verbatim}
\usepackage{graphicx}
\usepackage{cite}
\usepackage{siunitx}

\usepackage[nolist,nohyperlinks]{acronym}
\usepackage[noabbrev,sort&compress]{cleveref}
\usepackage{caption}
\usepackage{tabularx}
\usepackage{booktabs}
\usepackage{authblk}

\begin{acronym}
    \acro{RQ}{Research Question}
    \acro{UPM}{\emph{Universidad Politécnica de Madrid}}
    \acro{ETSISI}{\emph{Escuela Técnica Superior de Ingeniería de Sistemas Informáticos}}
    \acro{SQL}{Structured Query Language}
    \acro{ABS}{Acrylonitrile Butadiene Styrene}
    \acro{PLA}{Polylactic Acid}
    \acro{PETG}{Polyethylene terephthalate}
    \acro{TPU}{Thermoplastic polyurethane}
    \acro{LMS}{Learning Management System}
    \acro{MOOC}{Massive Online Open Course}
    \acro{CNC}{Computerized Numeric Control}
\end{acronym}

\begin{document}

\title{Using 3D printed badges to improve student performance and reduce dropout rates in STEM higher education}

\author[1,2]{Raúl Lara-Cabrera}
\author[1,2]{Fernando Ortega}
\author[1]{Edgar Talavera}
\author[1]{Daniel López-Fernández}

\affil[1]{Departamento de Sistemas Informaticos, Universidad Politecnica de Madrid, Madrid, Spain}
\affil[2]{KNODIS Research Group, Universidad Politecnica de Madrid, Madrid, Spain}

\date{}

\maketitle

\begin{abstract}
Students' perception of excessive difficulty in STEM degrees lowers their motivation and therefore affects their performance. According to prior research, the use of gamification techniques promote engagement, motivation and fun when learning. Badges, which are a distinction that is given as a reward to students, are a well-known gamification tool. This contribution hypothesizes that the use of badges, both physical and virtual, improves student performance and reduces dropout rates. To verify that hypothesis, a case study involving 99 students enrolled in a Databases course of computer engineering degrees was conducted. The results show that the usage of badges improves student performance and reduces dropout rates. However, negligible differences were found between the use of different kind of badges.
\end{abstract}

\section{Introduction}\label{sec:intro}

Heducation in STEM is critical to the development of society. However, for a long time now many students perceive it as difficult and the dropout rate in these areas is worryingly high~\cite{Araque2009Nov,Pappas2016Jul}. Furthermore, the health situation resulting from COVID-19 has further exacerbated this perception, negatively influencing student performance and motivation~\cite{Daniel2020Oct}. Therefore, it is necessary to use methodologies and techniques that promote learning in a fun, engaging, and motivating way. 

Previous research showed that the use of gamification techniques, and in general the game-based learning methodology, meets these needs and contributes to the improvement of student performance and motivation in STEM areas~\cite{literaturereviewbadges, analysisbadges,badgesformotivation,Lopez-Fernandez2021Mar,Gordillo2022Jan,Mayor2021Sep,Gordillo2020Dec}. Gamification consists of using the mechanics, aesthetics and reasoning of games to engage people, motivate activity, promote learning, and solve problems. While the best-known tools of this learning approach are badges, points, rewards, and leaderboards, gamification goes much further with activities such as contests, storytelling, character visualization, and problem solving.

This work focuses on badges, which are a distinction that is given as a reward to students. Many \acp{LMS} allow the creation and management of virtual badges. The main hypothesis of this work is that the use of badges, both physical and virtual, improves student performance. In this study, virtual badges are used and, in addition, the emerging technology of 3D printing to de-virtualize badges and turn them into something physical is employed. The choice of 3D printing over other alternatives, such us injecting modeling or \ac{CNC}, is due to its user-friendly design process and the significantly reduced production costs. The versatility of 3D printing allows to quickly and efficiently create unique badge designs with intricate details at a fraction of the cost of traditional manufacturing methods.

To validate this hypothesis, a case study involving 99 students enrolled in a Databases course of computer engineering degrees offered by \ac{ETSISI} of \ac{UPM} in Spain has been conducted. The design of the study is quasi-experimental, including one control group (without badges) and two experimental groups (one with virtual badges and the other with physical badges). In the experimental groups, students who achieved a grade greater than or equal to 9 out of 10 on the online quizzes carried out for each topic of the course could earn badges. The badges were designed to be easily identifiable with the corresponding topic.

The academic performance of the three groups of students were compared using two instruments: a pre-test to determine the initial level of knowledge of each group, and a post-test to determine the level of knowledge acquired in each group. The results show positive effects of using badges in terms of performance and dropout rates, but it seems that the type of badge used has no influence on these terms. 

The rest of the paper is structured as follows: \Cref{sec:related-work} studies other works related to the use of gamification techniques to improve student academic performance and motivation, with special emphasis on technical degrees. \Cref{sec:3d-printing} presents an introduction to 3D printing, the technology that has been used to transform badges awarded to students into physical objects, and a description of the printer, materials, and models used. \Cref{sec:research-methodology} and \cref{sec:results} present respectively the research methodology and experimental results used to empirically validate the impact of the use of physical and virtual badges. \Cref{sec:discussion} delves into the meaning, importance and relevance of the experimental results obtained, and \cref{sec:conclusion} states the answer to our research questions and makes recommendations for future work on the topic.

\section{Related work}\label{sec:related-work}

The impact of gamification in education to capture students' attention during teaching is a research area that has been extensively studied \cite{surveyimpactgami, surveyGami, surveyGami2}. In fact, gamification has been applied at various levels of education, from elementary education \cite{gameElementary}, through secondary education \cite{gameSecondary}, and higher education \cite{gamihigher}, even in areas as diverse as science \cite{surveyGamiScience} or organizational behavior \cite{gamiOrganizationalBehavior}.

There are multiple gamification techniques that allow the incorporation of this learning approach. One of the methods that attracts more attention is the incorporation of rewards to motivate learning \cite{rewardsCOVID, rewardsMarketing, rewards&ML}. Rewards such as money, badges, or praise within gamification are also known as extrinsic motivation \cite{extrinsic, pointrewards}. Extrinsic motivation can be easily reinforced when applied in interactive activities, although the impact on each individual depends on the concrete learning approach and the predominant learning styles of each student.

The use of rewards in educational environments where gamification techniques are applied is a widely studied area. Studies such as \cite{literaturereviewbadges, analysisbadges, badgesformotivation} demonstrate its positive impact on teaching and other studies such as \cite{moocs} analyze the usage of challenges, badges, and leaderboards in \acp{MOOC}. Other studies such as \cite{physicvsvirtual} conclude that the use of badges seems beneficial to increase users’ engagement and motivation. Studies such as \cite{badges&XP} go a step further by including more innovative techniques, such as including the delivery of experience points along with badges and checking their correlation with student learning in order to investigate the influence of these gamification elements. However, there are other studies such as \cite{Kyewski} showing that badges have less impact on motivation and performance than is commonly assumed.

Furthermore, some specific studies about the usage of badges can be found in computer engineering education. For example, \cite{cstwoexperiments} analyzes the impact of reward delivery through two experiments, the first with reward delivery and the second with heatmap visualizations that show a prediction of the student's success. Similarly, \cite{sofwareengeneering} shows that badge rewards are positively received by software engineering students. Other studies such as \cite{Jurgelaitis} go further, 
and suggest that the use of gamification techniques, including badges, can improve motivation and performance of computer engineering students.

Overall, the existing literature suggests that the integration of rewards in gamification techniques can positively impact student motivation. Nevertheless, despite the positive effects on motivation observed in several studies, literature reviews such as \cite{SurveyMedical, surveyGamificacion, surveyGamificacion2}, do not provide a conclusive resolution regarding the extent to which the use of rewards can enhance student performance, despite its demonstrated impact on motivation. In the same vein, a literature review framed in the computer engineering field \cite{SurveyGamificationSE} suggests that gamification techniques, including badges, can improve student engagement and, to a lesser extent, the student knowledge too. Nevertheless, that review concludes that there is still scarce scientific evidence on the subject as this research area is still in its infancy. Therefore, more research is needed to examine the use of badges from a purely performance standpoint.

To the best of the authors' knowledge, to date there is no study in the field of computer engineering education that analyzes and compares with a quasi-experimental approach the impact of badges, both physical and virtual, on the student performance and the dropout rate. The present contribution attempts to fill this gap.

\section{3D printing}\label{sec:3d-printing}

3D printing has revolutionized both industrial parts manufacturing and the accessibility of the process itself, allowing anyone with a 3D printer to create parts at home~\cite{bull2009democratization}. These devices melt different materials and deposit them layer by layer using a movable extruder nozzle. The choice of material depends on the part's specific characteristics. The most commonly used 3D printing materials, described in \Cref{tab:materials}, are affordable and widely available.

\begin{table}[!t]
\caption{Most commonly used 3D printed materials}\label{tab:materials}%
\centering
\begin{tabular}{|c|c|c|c|}
\hline
Name       & Properties                      & Durability & Difficulty to print\\
\hline
\acs{PLA}  & Easy to print, biodegradable    & Medium     & Low  \\
\hline
\acs{ABS}  & Durable and impact resistant.   & High       & Medium  \\
\hline
\acs{PETG} & Slightly flexible, durable      & High       & Medium  \\
\hline
\acs{TPU}  & Extremely flexible, rubber like & Medium     & High\\
\hline
\end{tabular}
\end{table}

The design and modeling of the parts are done in industrial and/or 3D modeling software such as Autodesk Fusion, Blender, and OpenSCAD, to name a few examples. Around these designs, communities have been forged to share, rate, and download the design of parts to be manufactured with 3D printers, Thingiverse\footnote{\url{https://www.thingiverse.com/}} being the best known of all them.

To convert the 3D model into a set of numerical control instructions that can be interpreted by printers, algorithms called slicers are used, the best known being cura\footnote{\url{https://ultimaker.com/software/ultimaker-cura}} and slic3r\footnote{\url{https://slic3r.org/}}. These algorithms fillet the 3D model into stacked layers whose thickness will determine the resolution and final appearance of the printed part. The most common thicknesses are currently in the range of \SI{0.1}{\milli\metre}-\SI{0.8}{\milli\metre}.

\begin{figure}
    \centering
    \includegraphics[width=0.4\textwidth]{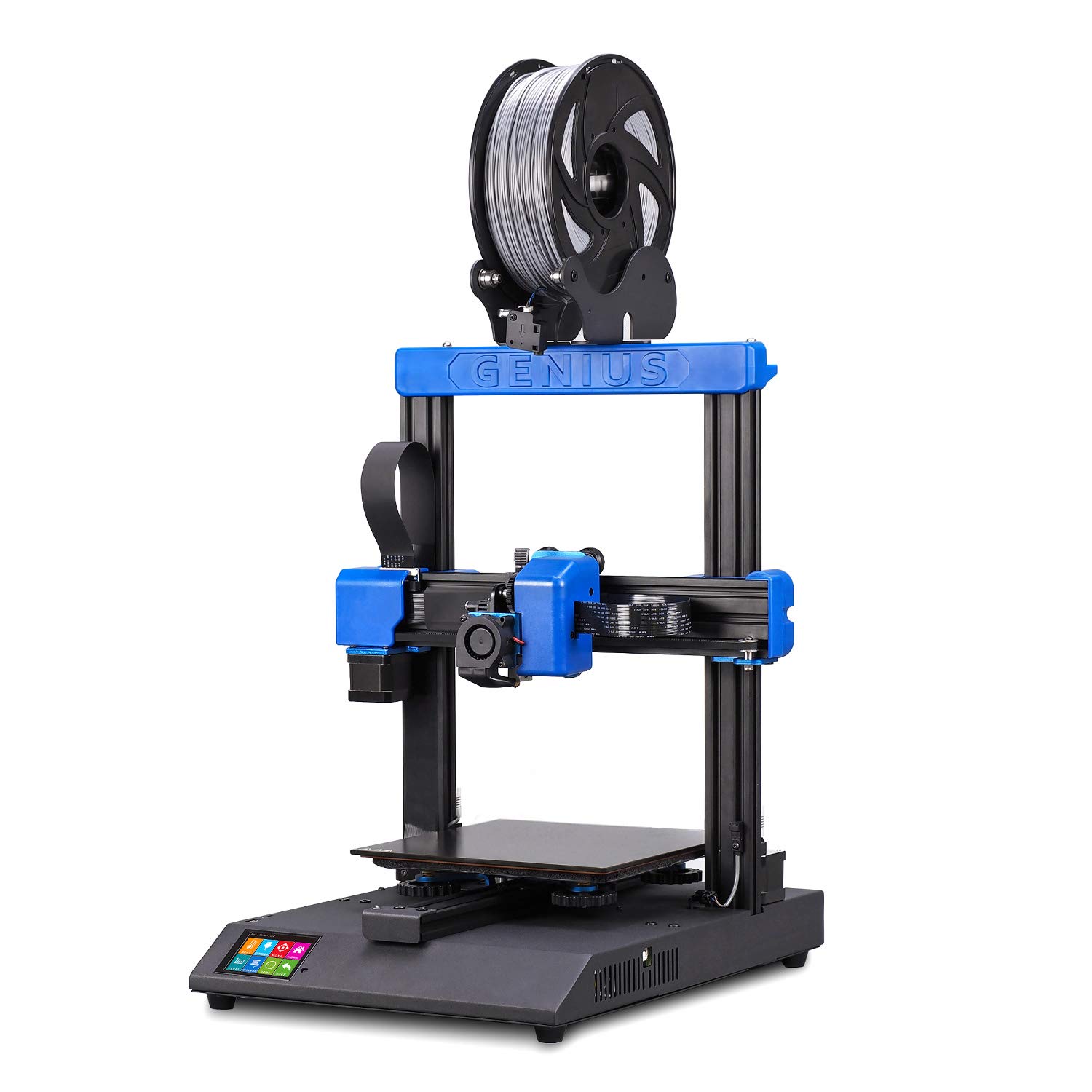}
    \caption{Artillery Genius, the 3D printer used in the experiment}
    \label{fig:genius}
\end{figure}

The printer used to print the physical badges delivered in the realization of the experiment was the Artillery Genius (see~\Cref{fig:genius}), a well-known budget 3D printer. Its key technical specifications are: cartesian unit with step-by-step driver, \SI{220x220x250}{\milli\metre} printing volume, \SI{150}{\milli\metre/\second} max printing velocity, AC hotbed with fast heating, \SI{0.05}{\milli\metre} minimum layer resolution, and  titan style direct drive extruder to support flexible materials.

The parts have been designed and modeled using Autodesk Tinkercad\footnote{\url{https://www.tinkercad.com/}}. Furthermore, badge models have been published in the Thingiverse design exchange community\footnote{\url{https://www.thingiverse.com/thing:5150131}}, so that any teacher can download and print them to use the badges in their classes. The printing material has been \ac{PLA} of different colors.

\section{Research methodology}\label{sec:research-methodology}

This section presents the research methodology used to empirically validate the impact of using badges, both physical and virtual, in STEM higher education.

\subsection{Research questions} \label{subsec:research-questions}

This research is designed to provide answers to the following \acp{RQ}:

\begin{itemize}
    \item RQ1: Does the use of badges influence the performance of computer engineering students?
    \item RQ2: Does the format of badges, virtual or physical, influence the performance of computer engineering students?
    \item RQ3: Does the use of badges influence the dropout rate of computer engineering students?
    \item RQ4: Does the format of badges, virtual or physical, influence the dropout rate of computer engineering students?
\end{itemize}

\subsection{Context and sample description}

This research was carried out at \ac{UPM}. \Ac{UPM} is one of the most prestigious universities in Spain with almost 40,000 students and more than 20 schools. \Ac{UPM} is mainly focused on engineering education. This study was carried out in the school named \ac{ETSISI}. This school offers four degrees related to computer engineering: Software Engineering, Computer Engineering, Information Systems, and Information Society Technologies. 

The case study was conducted during the academic year 2021-2022 in the context of the aforementioned degrees. Specifically, this research was done on the Databases course, which is a third semester course taught jointly for all \ac{ETSISI} degrees. The course addresses the conceptual, logical, and physical design of a database, the use of \ac{SQL} to create, read, update, and delete information in a relational database, the communication with the database in the Java programming language, and the utilization of CSV, JSON and XML files as an alternative to the databases.

The sample of this study was the students enrolled on the Databases course in the academic year 2021-2022. However, students who took the course were excluded as they had prior existing database knowledge. Thus, 99 students participated in this study.

In order to be able to validate the \acp{RQ} of \cref{subsec:research-questions}, the students in the sample were separated into 3 groups: students who will not receive any kind of badge to be used as \textbf{control group} (hereinafter referred as \emph{none} group), students who will receive virtual badges (hereinafter referred as \emph{virtual} group), and students who will receive physical 3d printed badges (hereinafter referred as \emph{3d printed} group). To perform this split, the academic organization at \ac{ETSISI} was taken into account. As mentioned above, the Databases course is a common course for the four degrees taught at \ac{ETSISI} and students are evenly grouped according to their degree and shift: Software Engineering students are grouped with Information Systems students (\emph{SE-IS} groups), and Computer Engineering students are grouped with Information Society Technologies students (\emph{CE-IST} groups). There exist 3 morning academic groups for \emph{SE-IS} students (\emph{SE-IS\_M1}, \emph{SE-IS\_M2} and \emph{SE-IS\_M3}), 2 morning academic groups for \emph{CE-IST} students (\emph{CE-IST\_M1} and \emph{CE-IST\_M2}), 2 evening academic groups for \emph{SE-IS} students (\emph{SE-IS\_E1} and \emph{SE-IS\_E2}), and 2 evening academic groups for \emph{CE-IST} students (\emph{CE-IST\_E1} and \emph{CE-IST\_E2}). \Cref{tab:groups-split} contains the groups to which each of the academic groups of the subject were assigned. This division was made by trying to ensure that all experimental groups had representation from all degrees and that all groups had students from the morning and evening turns. \Cref{tab:sample} shows the proportion of students assigned to each group. It is observed that all groups have approximately the same number of students.

\begin{table}[!t]
\caption{Groups to which each of the academic groups of the Databases course have been assigned.}
\label{tab:groups-split}
\begin{tabular}{l|ccccc|cccc|}
\cline{2-10}
                                 & \multicolumn{5}{c|}{SE-IS}                                                                                 & \multicolumn{4}{c|}{CE-IST}                                                      \\ \cline{2-10} 
                                 & \multicolumn{1}{c|}{M1} & \multicolumn{1}{c|}{M2} & \multicolumn{1}{c|}{M3} & \multicolumn{1}{c|}{E1} & E2 & \multicolumn{1}{c|}{M1} & \multicolumn{1}{c|}{M2} & \multicolumn{1}{c|}{E1} & E2 \\ \hline
\multicolumn{1}{|l|}{None}       & \multicolumn{1}{c|}{}   & \multicolumn{1}{c|}{X}  & \multicolumn{1}{c|}{}   & \multicolumn{1}{c|}{X}  &    & \multicolumn{1}{c|}{}   & \multicolumn{1}{c|}{}   & \multicolumn{1}{c|}{X}  &    \\ \hline
\multicolumn{1}{|l|}{Virtual}    & \multicolumn{1}{c|}{X}  & \multicolumn{1}{c|}{}   & \multicolumn{1}{c|}{}   & \multicolumn{1}{c|}{}   & X  & \multicolumn{1}{c|}{}   & \multicolumn{1}{c|}{X}  & \multicolumn{1}{c|}{}   &    \\ \hline
\multicolumn{1}{|l|}{3D printed} & \multicolumn{1}{c|}{}   & \multicolumn{1}{c|}{}   & \multicolumn{1}{c|}{X}  & \multicolumn{1}{c|}{}   &    & \multicolumn{1}{c|}{X}  & \multicolumn{1}{c|}{}   & \multicolumn{1}{c|}{}   & X  \\ \hline
\end{tabular}
\end{table}

\begin{table}[!t]
\caption{Number of students in each group.}
\label{tab:sample}
\centering
\begin{tabular}{|l|r|r|r|}
\hline
Group              & none & virtual & 3d printed \\ 
\hline
Number of students & 32   & 31      & 36         \\ 
\hline
\end{tabular}
\end{table}

\subsection{Methods and instruments}

As mentioned, this study seeks to motivate students to work steadily in the Databases course through rewards and reduce the dropout rate. During the Database course, students complete a total of 6 quizzes, one for each of the topics in the syllabus. These quizzes are carried out at the end of the topic, online and asynchronously by the students during the 12 hours they remain open. The main objective of the quizzes is that students perform a self-assessment on each of the topics of the course. This self-assessment allows them to measure their knowledge acquisition before taking the final exam of the course.

\begin{figure*}[!t]
    \centering
    \subfloat[Data modeling]{\includegraphics[width=0.25\textwidth]{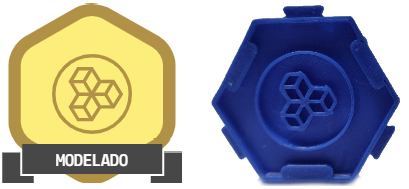}%
    \label{fig:virtual-badge-2}}
    \hspace{0.75cm}
    \subfloat[Relational model]{\includegraphics[width=0.25\textwidth]{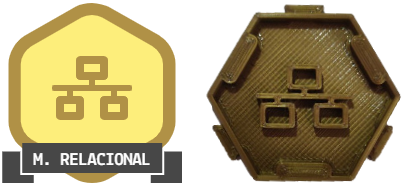}%
    \label{fig:virtual-badge-3}}
    \hspace{0.75cm}
    \subfloat[Structured Query Language]{\includegraphics[width=0.25\textwidth]{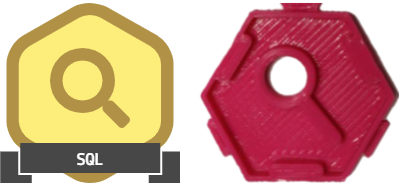}%
    \label{fig:virtual-badge-4}}

    \vspace{0.3cm}
    
    \subfloat[Databases management]{\includegraphics[width=0.25\textwidth]{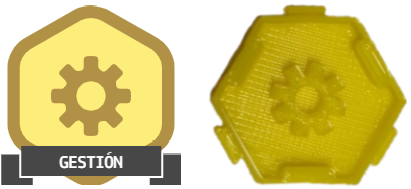}%
    \label{fig:virtual-badge-5}}
    \hspace{0.75cm}
    \subfloat[Programming with databases]{\includegraphics[width=0.25\textwidth]{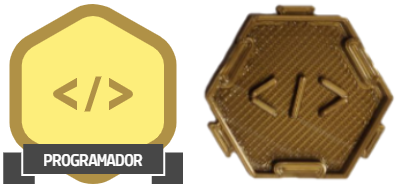}%
    \label{fig:virtual-badge-6}}
    \hspace{0.75cm}
    \subfloat[Files-based storage]{\includegraphics[width=0.25\textwidth]{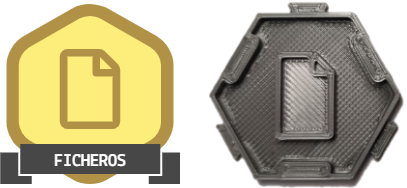}%
    \label{fig:virtual-badge-7}}
    
    \caption{Virtual and physical badges for each topic.}
    \label{fig:badges}
\end{figure*}

\begin{figure}[!t]
    \centering
    \includegraphics[angle=-3, width=0.35\textwidth]{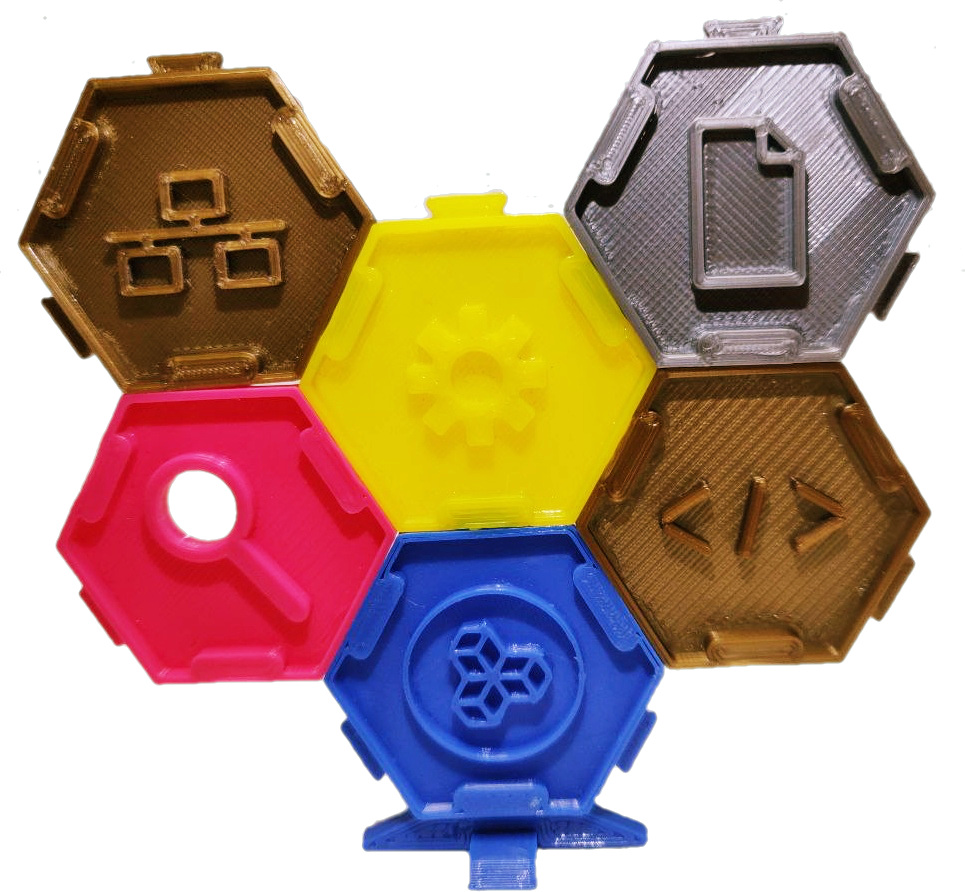}
    \caption{3D printed badges cluster}
    \label{fig:physical-badges-cluster}
\end{figure}

The teachers of the course know that the main cause of failure in the course is the absence of study hours of the students. Therefore, students are encouraged to take the quizzes and force themselves to study more and more regularly. In this vein, in order to extrinsically motivate students, when completing the quizzes, a badge will be awarded to those students who obtain a score equal to or higher than 9 out of 10 in the quiz. Two types of badges will be awarded: virtual badges through the Moodle \ac{LMS} platform and 3d printed physical badges, all of them shown in \cref{fig:badges}. The purpose of the physical badges is that they can be grouped on a cluster to be displayed on students' desks as shown in \cref{fig:physical-badges-cluster}.

The delivery of both physical and virtual badges takes place in the class immediately following the completion of the quiz. The badge award ceremony consists of: mentioning the students who have achieved the badge, awarding the virtual or physical badge, and encouraging the rest of the students to clap the badge achiever. The physical 3d printed badges were produced on demand: as soon as the results of the quizzes were available, the physical badges were printed. This was possible due to the reduced printing time of each badge (\SI{30}{\min} approximately).

Prior to the experience, during one of the lessons the first week of the course, a \textbf{pre-test} was carried out to determine the degree of knowledge of the students in the Databases course. The pre-test consisted of a test with 10 questions about the complete syllabus of the Databases course. This pre-test guarantees the absence of biases in the experiment groups shows on \cref{tab:groups-split}. The pre-test realization was supervised by a teacher.

Furthermore, a \textbf{post-test} was conducted to compare the academic results of the different experimental groups. The final exam of the course was used as a post-test. This exam consisted of a written examination with theoretical and practical questions on the entire course syllabus. The final exam lasts two hours and a half and takes place in university classrooms under the supervision of the teachers to ensure that no student cheats.

Moreover, the \textbf{dropout rate} during the academic year 2021-2022 was also measured, both in general and in each of the groups analyzed.

Finally, a \textbf{survey} was voluntary fulfilled by the students of the physical and virtual badge groups in order to assess the effect of badge use on their motivation. The questions asked to the students in were:

\begin{description}
\item[Q1]: My overall opinion about the use of badges is positive
\item[Q2]: The use of badges seems attractive and motivating to me
\item[Q3]: The use of badges helps to make learning more fun
\item[Q4]: The use of badges positively contributes to my learning process
\item[Q5]: I believe that the use of physical badges enhances the educational experience based on virtual badges
\item[Q6]: I would like other courses to also use badges, whether virtual or physical
\item[Pref.]: If you have used physical (virtual) badges, would you have liked to use them in virtual (physical) format?
\end{description}

\subsection{Data analysis techniques}

First of all the Shapiro–Wilk test was performed to check the normality of the data (i.e., survey, pretest, post-test and learning gains, which were calculated as the difference between post-test and pretest scores). As the data were normally distributed, parametric statistical methods were employed. 

The quantitative data obtained from the results of the survey, the pre-test, the grades of the students in the quizzes and the post-tests were analyzed in several ways: 
\begin{enumerate}
    \item Using \textbf{box-plot diagrams} to get a visual representation of the differences between the groups of this study.
    \item Using \textbf{histograms} to compare the distribution of the grades in the gropus of this study.
    \item Using \textbf{statistical tests} to ensure the validity of the experimental results. For this purpose, both ANOVA and ANCOVA tests were used. The ANOVA test is used to check if significant difference exists in the results of the pre-test and the learning gains. The ANCOVA test is used to deepen into the differences of the academic results that exist between the groups of this study.
    \item Using a \textbf{statistical test} to assess if there was a statistically significant difference between the survey results obtained by the physical and virtual badges groups. Specifically, Mann-Whitney tests were carried out.
\end{enumerate}

Lastly, the dropout rate was analyzed by using descriptive statistics.

\section{Results}\label{sec:results}

This section contains the results used to answer the research questions of this study empirically and reliably.

\subsection{Knowledge acquisition}
\begin{figure}[!t]%
\centering
\includegraphics[width=0.4\textwidth]{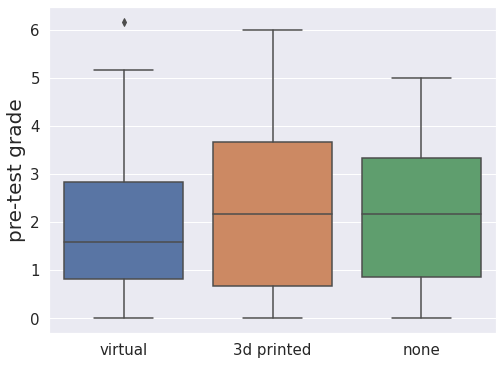}
\caption{Pre-test grade split by groups.}\label{fig:pre-test}
\end{figure}

The results of the pre-test are depicted in \cref{fig:pre-test}. As it can be observed, there are no major differences between the groups. On the one hand, the \emph{none} and \emph{3d printed} groups are very similar, although the latter has a slightly higher standard deviation. On the other hand, the \emph{virtual} group seems to have slightly worse results than the other groups. However, to check if there is a difference between the groups, an ANOVA test was performed. The result of this test was 0.2235 (p $>$ 0.05), so it can be stated that there is no significant difference between the groups in the pre-test.

\begin{figure}[!t]%
\centering
\includegraphics[width=0.4\textwidth]{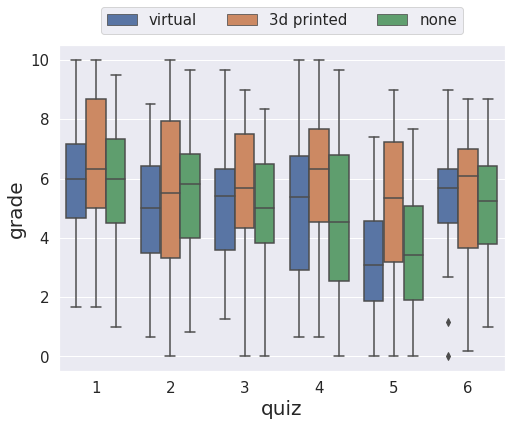}
\caption{Grades obtained in each of the quizzes split by groups.}\label{fig:quizzes}
\end{figure}

\Cref{fig:quizzes} shows the grades obtained in each of the quizzes divided by groups. The general trend is that both the median and the interquartile range of the \emph{3d printed} group are higher than those of the \emph{virtual} and \emph{none} groups, so it seems that students in the \emph{3d printed} group tend to get better grades.

\begin{figure}[!t]%
\centering
\includegraphics[width=0.4\textwidth]{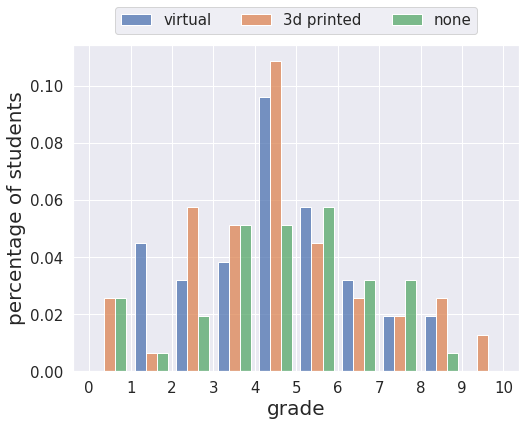}
\caption{Percentage histogram of the grades in this final exam split by groups.}\label{fig:post-test}
\end{figure}

Although these results seem to confirm the hypothesis that the use of badges improves the performance of STEM students, quizzes are not a valid test to measure knowledge acquisition because they are performed online without any supervision. Quizzes are a way for students to study, but they do not reliable assess the student's knowledge nor do they determine the final grade of the course. As stated above, the grade is determined by a final exam consisting of a written test that takes place in a university classroom under the supervision of the teacher, so the final exam is used as post-test. \Cref{fig:post-test} contains the percentage histogram of the grades in this final exam split by groups. It shows that the students in the badge groups have higher grades than the students in the \emph{none} group, with the majority of these students being in the \emph{3d printed} group.

\begin{figure}[!t]%
\centering
\includegraphics[width=0.4\textwidth]{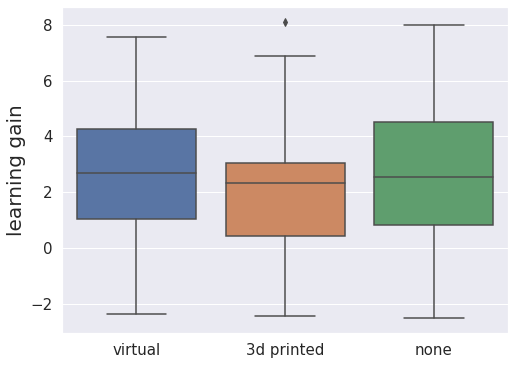}
\caption{Learning gains between the pre-test and post-test split by groups.}\label{fig:learning-gain}
\end{figure}

\Cref{fig:learning-gain} shows the learning gains (i.e., the difference between the post-test score with respect to the pre-test score) categorized by groups. It can be observed that: (a) the median in the three groups is the same, (b) the variance in the ratings of the \emph{3d printed} group is lower, and (c) the best grades occur as outliers in the \emph{3d printed} group. An ANOVA test for these data returns a p-value of 0.5926 (p $>$ 0.05), so, although they show some differences in the learning gains of the groups, they are not significant.

To deepen the analysis of the data and better understand the implication of badges in the learning gains attained by the students, two ANCOVA tests were performed: one comparing the usage of badges (virtual or 3d printed) vs. the not usage of badges, and another comparing the academic results of the 3 groups (\emph{none} group, \emph{virtual} group and \emph{3d printed} group). By examining the confidence intervals, a confident of the 95\% is obtained for the following statement:

\begin{enumerate}
    \item Using badges versus not using badges implies a difference in learning gains of between -1\% and 17\%.
    \item Using 3d badges versus using nothing implies a difference in learning gains of between -7\% and 17\%.
    \item Using 3d badges versus using virtual badges implies a difference in learning gains of between -18\% and 7\%.
    \item Using virtual badges versus using nothing implies a difference in learning gains of between -2\% and 23\%.
\end{enumerate}

In view of these results, it can be concluded that the use of physical or virtual badges is clearly preferable to not using badges at all.

\subsection{Dropout rate}

\begin{figure}[!t]%
\centering
\includegraphics[width=0.35\textwidth]{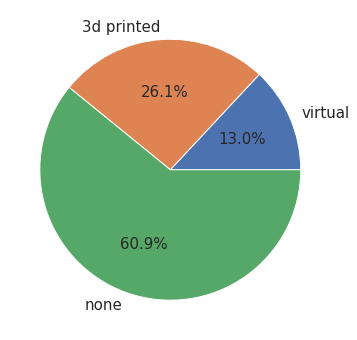}
\caption{The group to which the students who took the quizzes but did not take the final exam belonged.}\label{fig:dropout}
\end{figure}

In order to meet the objectives of this research, it is also important to analyze the dropout rate that occurs in each of the groups. \cref{tab:dropout-rates} shows the number and percentage of students in each group who dropped out of the course (i.e., did not take the final exam). On the other hand, \cref{fig:dropout} shows, with respect to the total number of dropouts, the percentage of students who belonged to the \emph{none} group, the \emph{virtual} group and the \emph{3d printed} group. It can be observed that the groups in which there was the possibility of receiving badges suffered much lower dropout rates.

\begin{table}[!t]
\caption{Number $N$ of students in each group, number $N_{\textrm{drop}}$ and percentage $R_{\textrm{drop}}$ of students in each group who dropped out of the course.}\label{tab:dropout-rates}
\centering
\begin{tabular}{|l|r|r|r|}
\hline
Group      & $N$ & $N_{\textrm{drop}}$ & $R_{\textrm{drop}}$ \\ \hline
none       & 32  & 14                  & 43.75\%             \\ \hline
virtual    & 31  & 3                   & 9.67\%              \\ \hline
3d printed & 36  & 6                   & 16.66\%             \\ \hline
total      & 99  & 23                  & 23.23\%             \\ \hline
\end{tabular}
\end{table}

\subsection{Students perceptions towards the use of badges}

\begin{figure}[!t]%
\centering
\includegraphics[width=\columnwidth]{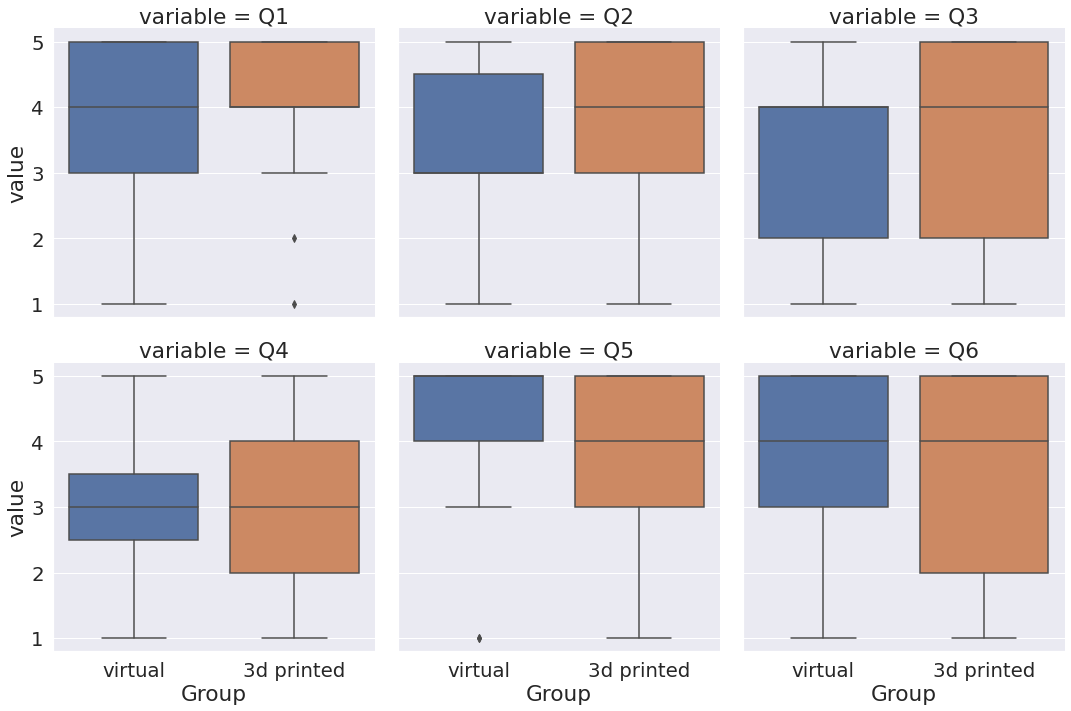}
\caption{Students' responses to each question asked in the voluntary surveys on physical ($n=17$) and virtual ($n=15$) badges.}\label{fig:survey-boxplots}
\end{figure}

\Cref{tab:survey} displays the results obtained in the voluntary surveys conducted among the students in the virtual badge and physical badge groups. The responses use a Likert scale with values ranging from 1 to 5, where 1 means 'strongly disagree' and 5 means 'strongly agree'. The results shown in the table corresponds to the mean value and standard deviation of the responses collected for each question. Moreover, \cref{fig:survey-boxplots} depicts graphically the students’ responses to each question asked in the survey. In view of the high ratings obtained, it can be appreciated that the use of both physical and virtual badges, is very well received by students (Q1, Q6) and they found this gamification technique attractive and motivating (Q2), fun (Q3) and helpful for their learning process (Q4).

\begin{table*}[ht]
\centering
\resizebox{\textwidth}{!}{%
\begin{tabular}{l|llllll|l}
\toprule                & Q1           & Q2            & Q3            & Q4            & Q5            & Q6 & Preference           \\ \midrule
Physical badges & $4.06 \pm 1.20$ & $3.94 \pm 1.30$  & $3.41 \pm 1.42$ & $2.88 \pm 1.22$ & $3.82 \pm 1.33$ & $3.24 \pm 1.60$ & YES: 18\%, NO: 47\% \\
Virtual badges  & $3.80 \pm 1.37$ & $3.33 \pm 1.35$ & $3.27 \pm 1.39$ & $3.00 \pm 1.36$    & $4.13 \pm 1.41$ & $3.73 \pm 1.39$ & YES: 80\%, NO: 0\% \\ \bottomrule
\end{tabular}%
}
\caption{Results obtained (mean and standard deviation on a Likert scale from 0 to 5) for each question asked in the voluntary surveys on physical ($n=17$) and virtual ($n=15$) badges.}\label{tab:survey}
\end{table*}

\begin{table}
\centering
\begin{tabular}[t]{r|r|r|r|r}
\toprule
Variable & $n_{virtual}$ & $n_{physical}$ & Statistic & p-value\\
\midrule
Q1 & 15 & 17 & 116.5 & 0.673\\
Q2 & 15 & 17 & 93.0 & 0.182\\
Q3 & 15 & 17 & 119.0 & 0.757\\
Q4 & 15 & 17 & 131.5 & 0.891\\
Q5 & 15 & 17 & 147.0 & 0.434\\
Q6 & 15 & 17 & 149.0 & 0.412\\
\bottomrule
\end{tabular}
\caption{Mann-Whitney tests results}\label{tbl:mann}
\end{table}

In order to deepen into the differences obtained between both groups, Mann-Whitney tests were carried out and the results are reported in (see~\cref{tbl:mann}). Despite the lack of statistical significance, the obtained differences and, specially the difference in percentage in the preference variable, suggest that students prefer physical badges. 

\section{Discussion}\label{sec:discussion}

The results allow us to answer the research questions presented above (Section \ref{subsec:research-questions}). 

Regarding RQ1 (\textit{Does the use of badges influence the performance of computer engineering students?}), the results of the online quizzes and the analysis of learning gains suggest that the use of badges improves student performance because the students of the badge groups obtained higher scores than the students of the group who did not use badges. Although the ANOVA test does not reveal significant statistical differences in this respect, the results from the ANCOVA test reveal that the use of rewards implies a positive difference (up to +17\%) in favor of the groups in which badges were used. These results are contrary to the ones reported in \cite{Kyewski}, and partially consistent with some studies conducted with virtual badges \cite{reward3D, gamihigher} and physical badges \cite{physicvsvirtual}, which conclude that positive results in terms of student performance occur when using badges but they are not extremely significant. However, our results are consistent with other many studies such as \cite{Jurgelaitis,cstwoexperiments, sofwareengeneering, badgesformotivation, rewards&ML} and the body of scientific research on gamification reported by several literature reviews \cite{SurveyMedical, surveyGamificacion, surveyGamificacion2,SurveyGamificationSE,dropout}, which in general terms suggest that a significant improvement in student outcomes can occur when gamification techniques such as badges are used.

Regarding RQ2 (\textit{Does the format of badges, virtual or physical, influence the performance of computer engineering students?}), the knowledge acquisition results suggest that the type of badges, virtual or physical, hardly affect student performance. The use of 3d badges versus the use of virtual badges implies a difference in learning gains between -18\% and 7\%, therefore in overall terms there is a slight difference in favor of the virtual badges. Anyhow, this difference is so small that we can conclude that the format of the badges does not influence student performance. However, this is only true for the badge designs presented in this study. If the design of the 3D printed badges were different this could change. The conclusions regarding this research question are one of the novelties of the study, because to the authors' knowledge there are no similar studies comparing the effect of virtual badges with 3D-printed badges in computer engineering education.

Regarding RQ3 (\textit{Does the use of badges influence the dropout rate of computer engineering students?}), the analysis performed with the dropout rates clearly indicates that the use of badges contributes positively to a decrease in the dropout rate. The percentage of students who drop out of the course in the group where badges were not used is over 40\%, while in the other groups it is between 9\% and 16\%. Although the analyzed related work does not analyze this badge effect, some literature reviews such as \cite{dropout} suggest that this change is a positive side effect obtained by improving student motivation. This is the premier result of this research because, to the best knowledge of the authors, it shows for the first time the positive effect of badges on dropout in computer engineering education, which is a critical problem in STEM education~\cite{Araque2009Nov,Pappas2016Jul}.

Regarding RQ4 (\textit{Does the format of badges, virtual or physical, influence the dropout rate of computer engineering students?}), the aforementioned analysis with dropout rates indicates that the use of virtual badges contributes slightly more positively to the decrease in absenteeism than the use of physical badges. Specifically, the percentage of students who dropout of the course in the virtual group was 9.97\%, while in the 3d group was 16.66\%. Anyhow, as in the case of RQ2, the difference between both groups is so small that we can conclude that the format of the badges does not seem to influence student dropout (at least considering the design of the 3D printed badges presented in this paper). Once again, the conclusions regarding this research question are one of the novelties of the study, because to the authors' knowledge there are no similar studies comparing the effect of virtual badges with 3D-printed badges in computer engineering education.

Furthermore, to continue discussing the posed research questions it is useful to comment the results obtained in the survey since they allow to better understand the impact of the badges from a motivational point of view, as well as to compare the effect of virtual and physical badges from this perspective. One of the factors contributing to the positive impact of badges on performance and dropout rate is the fact that badges, both virtual and physicals, are attractive, motivating, and fun for the students. Indeed, they claimed that this gamification technique was helpful for their learning process. The results comparing the motivational impact of of virtual and physical badges did not yield a statistically significant difference, which is is consistent with the performance and dropout rate results. However, caution should be exercised in interpreting the survey results as the small sample size may limit the reliability and generalizability of these findings. 
Finally, in order to extend the discussion about the motivational impact of the badges initiated from the empirically obtained data, a detailed analysis of the motivational mechanisms influencing student performance is presented. To do so, the research presented in \cite{lopez2015motivation} is of great interest because it analyzes in depth several motivational indicators that affect engineering students. These indicators are based on well-known motivational theories such as the needs theory from McClelland, the equity theory from Adams, the expectations theory from Vroom, or the dual factor theory from Herzberg. Following these theories the usage of badges are clearly useful because it enhances several motivational indicators:
\begin{enumerate}
    \item Achievement (Needs theory, McClelland): For a student, being able to achieve a certain score on a test and thus earn a badge fosters achievement motivation, which brings personal satisfaction.
    \item Power (Needs theory, McClelland): When a teacher awards a badge to a student, it is public recognition of the student's performance that can fuel the power motivation, which is essential for many people.
    \item Affiliation (Needs theory, McClelland): If several students in a group have earned a badge, it is quite possible that those students who have not yet received a badge will make an effort to earn the badge themselves, thus fostering the need to belong to the group.
    \item Effort/reward (Equity theory, Adams): Badges are an additional reward for a student, who make an effort to prepare for the subject. In this way, badges are a mechanism to make the reward side of the effort/reward balance weigh a little more. 
    \item Desire (Expectations theory, Vroom): For many students, earning a badge is important due to many reasons like those mentioned above (achievement, power, affiliation, etc.). This desire can be an additional motivation to study.
    \item Reward (Dual theory, Herzberg): In the same vein as effort/reward indicator, badges are a motivator for students. Examining this through the prism of Herzberg's theory, it is important to note that badges are an extrinsic motivator. While it is certainly true that a student should foster his/her intrinsic motivators (e.g., enjoyment of study, sense of intellectual progress, etc.), extrinsic motivators that a teacher may offer, such as badges, are a positive element in the teaching-learning process that can be unleashed to foster intrinsic motivation as well.
\end{enumerate}

This analysis from the motivational point of view allows a better understanding of the results obtained since it offers theoretical explanations to the empirical results that support the answers to the posed research questions. In relation to RQ1 and RQ3, the analyzed theories allow to infer that badges are a mechanism to increase student motivation from many perspectives (e.g., need for achievement, power or affiliation, obtaining rewards, etc.), which ends up influencing their performance. In relation to RQ2 and RQ4, none of the analyzed theories distinguish between the format of the rewards since they analyze what the reward implies (e.g., personal satisfaction, public recognition, etc.). Therefore, for these theories, the fact that a student receives a reward, whether digital or physical, is already motivating as shown by the results obtained empirically in this research.

\section{Conclusion}\label{sec:conclusion}

This paper presents a rigorous quasi-experimental study to examine how the delivery of virtual and physical badges can affect student performance in the field of computer engineering. The results show that both kinds of badges clearly improve student performance with respect to the non-use of badges (i.e., the students who receive badges obtained better academic results than the students of the control group). Moreover, the differences between the outcomes produced by virtual badges and 3D printed badges are minimal (i.e., the students who receive virtual badges obtained academic results similar to the students that receive 3D printed badges).

Regarding the other variable under examination, this article studies the influence of badges on the dropout rate of computer engineering students. The obtained results indicate that the dropout rate when badges are used as a motivational method is around 10-15\%, while the dropout rate when no badges are used is over 40\%. So, the use of badges has a very positive effect on the dropout rate. This is the premier result of this research, so it shows for the first time the positive effect of badges on dropout in computer engineering education. Moreover, by comparing the dropout rate of the group who received virtual badges and the group who received 3D printed badges we found negligible differences and we can conclude that the format of the badges does not seem to influence the student dropout rate.

However, the conclusions obtained in this study and especially those related to the comparison between the academic impact of virtual and physical badges are tied to the type of design of the 3D printed badges. If more appealing  badges were designed, different results could be obtained. In fact, future work involve the replication of this experiment but using different types of badges and the study of the influence that the badge design itself could have on the student performance.

Nevertheless, in view of the results of this experiment, what seems clear is that regardless of the design of the badges, the use of this gamification technique can improve academic performance and reduce the dropout rate of computer engineering students. To consolidate this conclusion it would be interesting  to replicate the presented experiment in more computer engineering courses. Additionally, another interesting idea to work on the future is to explore competition among students with different scales of rewards according to the results obtained. 

\section*{Acknowledgments}
This work was supported by the \textit{Comunidad de Madrid} under \textit{Convenio Plurianual} with the \textit{Universidad Politécnica de Madrid} in the actuation line of \textit{Programa de Excelencia para el Profesorado Universitario}.

\bibliographystyle{plain}
\bibliography{bibliography}

\end{document}